# Plasmon-Assisted Suppression of Surface Trap States and Enhanced Band-Edge Emission in a Bare CdTe Quantum Dot

*Assegid M. Flatae,\*[a] Francesco Tantussi,[b] Gabriele C. Messina, [b] Francesco De Angelis[b], and Mario Agio[a,c]*

[a]  Laboratory of Nano-Optics and Cµ

   University of Siegen

   57072 Siegen, Germany

   \* flatae@physik.uni-siegen.de

[b]  Istituto Italiano di Tecnologia

   16163 Genova, Italy

[c]  National Institute of Optics (INO-CNR)

    50125 Florence, Italy

**Abstract:** Colloidal quantum dots have emerged as a versatile photoluminescent and optoelectronic material. Limitations like fluorescence intermittency, non-radiative Auger recombination and surface traps are commonly addressed by growing a wide-bandgap shell. However, the shell isolates the excitonic wave function and reduces its interaction with the external environment necessary for different applications. Furthermore, their long emission lifetime hinders their use in high-speed optoelectronics. Here, we demonstrate a high degree of control on the photophysics of a bare core CdTe quantum dot solely by plasmon-coupling, showing that more than 99% of the surface defect-state emission from a trap-rich quantum dot can be quenched. Moreover, the band-edge state excitonic and biexcitonic emission rates are Purcell enhanced by 1460 and 613-fold, respectively. Our findings show how plasmon-coupling on bare quantum dots could make chemical approaches developed for improving their optical properties unnecessary, with implications for nanoscale lasers, light emitting devices, solar cells, and ultrafast single-photon sources.

Colloidal semiconductor quantum dots (QDs) represent a promising nanoscale material for application in optoelectronics, photovoltaics and in the life sciences. The size-tunable electronic structure, cheap solution processing, high photostability at room temperature and high fluorescence quantum yield has led to their integration into devices like light emitting diodes, lasers, solar cells, photodetectors, field-effect transistors and memory elements [1]. However, colloidal QDs suffer from fluorescence intermittency (on/off blinking), Auger recombination and surface traps. To minimize these effects, QDs are typically coated by a ligand or a wide-bandgap shell. QDs with thicker shells (core/shell QDs) are known to exhibit improved photoluminescence quantum yield of band-edge emission [2-4]. On the other hand, wide-bandgap shells isolate the QDs from the outside environment by confining the wave function of an exciton (electron and hole pair), limiting different types of applications [1]. For example, QD based light-emitting diodes would require efficient charge injection into the QDs [5-7]. Hence, an ideal core/shell QD used for optoelectronic applications should avoid intrinsic traps and be sensitive to external carriers with a certain level of drive [8]. Synthesis of such kind of core/shell QDs without trap states for optoelectronic applications is complex. Furthermore, a concrete knowledge on the trapping efficiency as a function of the shell-thickness is required, which so far remains elusive [9-13]. Furthermore, the long emission lifetime of both core and core/shell QDs (few tenth of ns) hinders their use in high-speed optoelectronics. Therefore, bare core QDs that have a suppressed surface trap state, an enhanced band-edge state emission, and a high radiative decay rate with less fluorescence intermittency are desirable.

Hybrid quantum systems based on plasmonic nanostructures offer outstanding possibilities to control the emission dynamics of a single emitter. Recent experiments have shown large fluorescence enhancement factors using plasmon coupling. The enhancement is attributed to the overall fluorescence yield obtained, including the effect of excitation enhancement and increase in quantum efficiency [14-16]. A thousand-fold change in the radiative decay rate was also reported [17,18], based on fluorescence lifetime measurements on ensembles of emitters. So far reports at the single-emitter level are rare [19-21] and none addresses bare QD systems. In this work we report on the complete suppression of trap-states emission in a strongly-confined bare-core single CdTe QD and on the enhanced fluorescence emission from its band-edge states by coupling the QD to a gold nanocone. Acting as nanoresonators, plasmonic nanocones provide a high degree of control on the photophysics of bare core CdTe QDs. Moreover, precise experimental determination of the quantum efficiency, radiative and non-radiative decay rates and the Purcell enhancement of excitonic and biexcitonic emissions of a single QD are presented.

We fabricate high-quality gold nanocones with dimensions in the 100 nm range using electron beam-induced deposition (EBID) of an organometallic precursor (containing platinum carbide) followed by sputtering deposition of a gold layer [22]. The nanocones exhibit a very sharp tip, with a radius of curvature down to 10 nm. The fabrication technique allows precise control over size, shape and radius of curvature, which is required for manipulating the photophysics at the single-emitter level. Figure 1(a) displays a scanning electron-micrograph image of a platinum-carbide bare nanocone and a gold-coated nanocone after sputtering deposition. The nanocones are designed to have resonances in the near-infrared region (NIR) to match the emission wavelength of CdTe QDs, as shown in Figure 1(b). The NIR region is also of interest because absorption in gold is smaller. Colloidal QDs have a size of ~ 5 nm (smaller than its excitonic Bohr radius ~7.3 nm) and are in the strong-confining regime. This results in well separated electronic states with respect to the available thermal energy leading to stable room-temperature emission properties. Here we choose a QD that emits at a central wavelength of 720 nm and that exhibits a slow excitonic and a relatively fast biexcitonic lifetime of 22 ns and 6 ns, respectively, as shown in Figure 1(c).

To optically access the QD from top and bottom, gold nanocones are fabricated on a transparent silicon-nitride membrane. A single CdTe QD approaches a nanocone by means of an atomic force microscope (AFM) probe from the top. QDs are attached to AFM probes by tip functionalization strategies. To prevent quenching of the emitter a silicon nitride tip is preferred to a gold-coated tip. Silanization of the AFM tip offers the advantage that the probe can be functionalized directly without other prior surface preparation. Amine groups (–NH2) are introduced to the tip surface via esterification by reaction of surface silanol groups with ethanolamine [HO-(CH$_2$)$_2$-NH$_2$]. The CdTe QDs that form colloidal solutions in water are terminated with a carboxyl (–COOH) group. This group is activated for direct conjugation to amine through carbodiimide (EDC)-mediated esterification. This enables carboxyl group terminated QDs to be attached to the amine-functionalized tip as shown in Figure 1(d). The dielectric effect of the AFM tip on the emitter is suppressed by using the appropriate ligand length. This offers the possibility of imaging the dipole moment orientation of a single QD, which is an important parameter as it dictates the coupling strength with the gold nanocone.

The QD attached to the AFM tip is brought to the near field of a gold nanocone in a controlled way by a closed-loop 6-axes piezo stage (Nanowizard®4, JPK Instruments AG). The system offers tip positioning and long-time position stability (position noise level < 0.06 nm RMS in the *xy*-plane and < 0.03 nm RMS along the *z*-axis). Analysis of AFM force curves helps to precisely determine the QD-nanocone distance with better than 1 nm resolution.

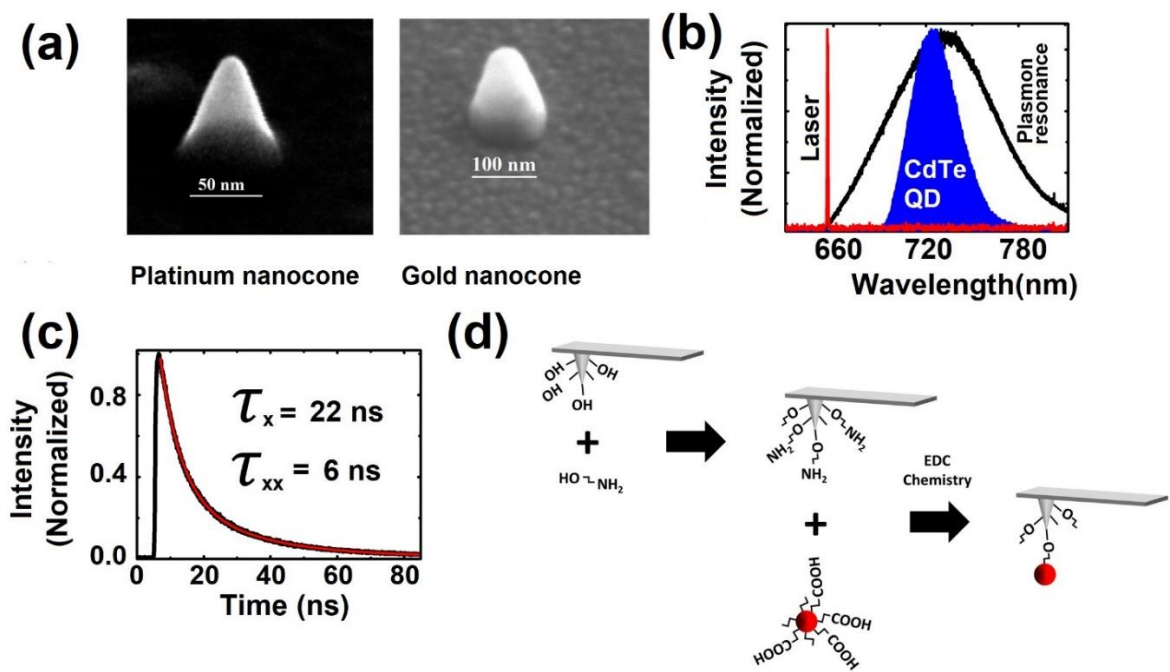

**Figure 1.** (a) Scanning electron micrograph image of a platinum-based nanocone (left) and a gold nanocone (right). (b) Spectral matching of a CdTe QD with a gold nanocone. The incident laser is detuned from resonance to avoid excitation enhancement. (c) The excited-state lifetime of a CdTe QD placed at the tip of a dielectric AFM probe. (d) Functionalization of silicon nitride AFM tips to attach QDs using EDC chemistry.

Individual coupled QD-nanocone hybrid systems are characterized using a home-built temporal- and spectrally-resolved confocal-microscopy (built on the base of an inverted microscope, Axio Observer 5, Carl Zeiss Microscopy GmbH) setup connected to a Hanbury-Brown Twiss (HBT) interferometer, as shown in Figure 2 (a). The sample is excited by a 656 nm cw/pulsed laser at a power < 1mW (PicoQuant, PDL 800-D, LDH-D-C-660) and the emission from the sample is collected via a cover-slip corrected oil-immersion microscope objective (Olympus, 60X, 1.42 NA, 0.15 mm working distance). P-polarized excitation at the location of the hybrid system is implemented to create an electric field component along the nanocone axis (Figure 2(b)). The spectrometer (Andor, Shamrock 500i) is equipped with an electron multiplying CCD (EMCCD) camera (Andor, Newton 970, A-DU970P-BVF) having a quantum efficiency around 90% at the emission wavelength of the QD. Imaging is performed using a flippable reflecting mirror that is placed along the optical axis of the setup. After wide-field laser illumination, the collected signal is sent to an EMCCD camera (Princeton Instruments, ProEM-HS: 512 BX3, back-illuminated EMCCD, more than 90% quantum efficiency at the emission wavelength of the QD). The anti-bunching behavior of a single QD is monitored using the second-order intensity autocorrelation function. The laser excites the hybrid system and the photoluminescence from the sample is spectrally filtered to suppress the laser. A 50/50 non-polarizing beam splitter sends the emitted photons to two avalanche photodiodes (APDs) (Micro Photon Devices, 50 cps dark count, < 50 ps jitter) as shown in Figure 2(c). These detectors are connected to the start-stop time-interval analyzer of a time-correlated single photon counter (TCSPC) (PicoQuant, PicoHarp 300) and the delay between the arrival times of two emitted photons is repeatedly measured and histogrammed with picosecond time resolution. To avoid cross talk between the two APDs, a band pass filter is placed in front of them. The excited-state lifetime is measured using the pulsed-mode operation of the diode laser. The overall instrument response function (IRF) of the setup is less than 70 ps.

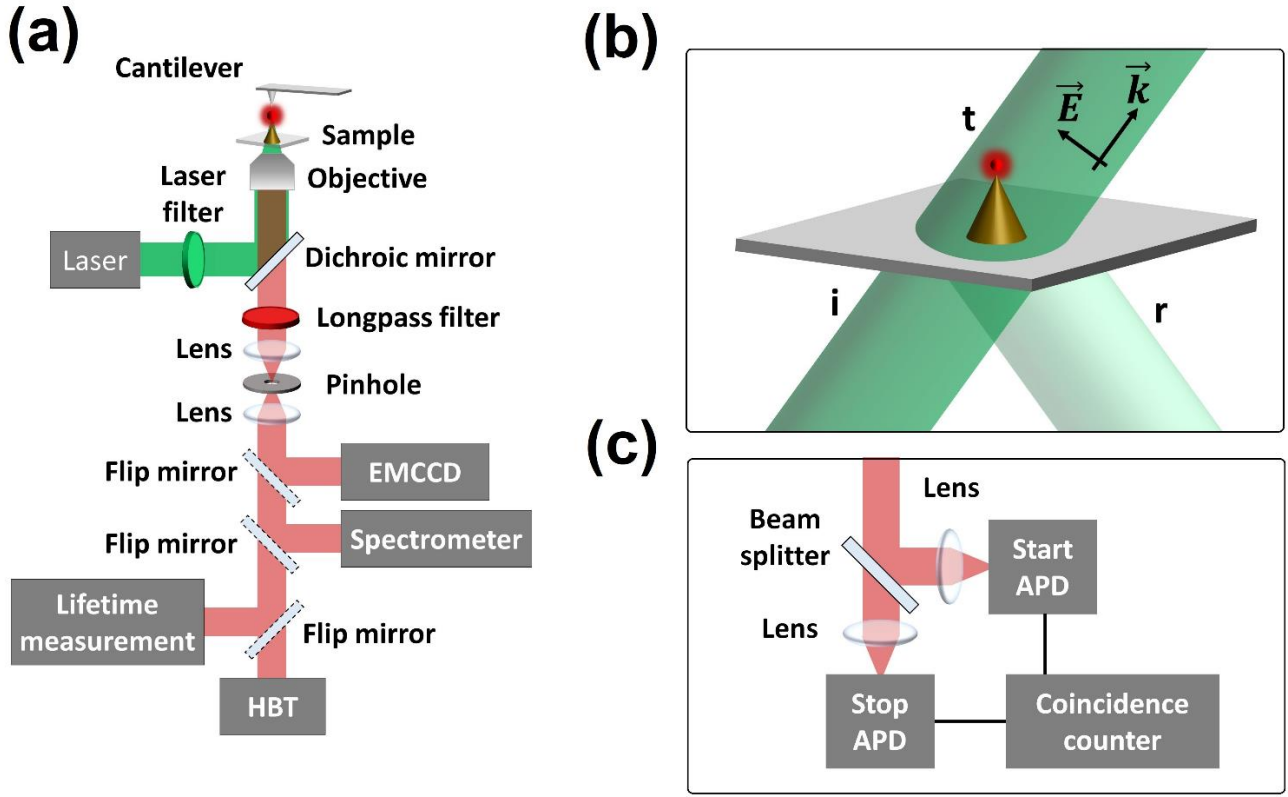

**Figure 2.** (a) Schematics of the experimental setup for imaging, photoluminescence, excited-state lifetime measurements and HBT interferometer. The QD attached to the tip of an AFM probe is brought to the nanocone with nm precision. (b)The incident p-polarized light (i) has a component along the nanocone axis. Part of the light is reflected (r) at the substrate, while the transmitted light (t) effectively excites the vertical dipole moment of the QD (c) A 50/50 non-polarizing beam splitter sends the emitted photons to two avalanche photodiodes (APDs). The delay between the arrival times of two emitted photons is repeatedly measured and histogrammed with picosecond time resolution.

Figure 3(a) shows the spectrum of a trap-rich single CdTe QD exhibiting both band-edge and trap-state emission. In the measured spectral window more than 74% of the fluorescence signal is obtained from surface-trap emission. The photon statistics is also monitored and it exhibits fluorescence intermittency for continuous excitation, as shown in Figure 3(b). This demonstrates the interrogation a single QD (also verified by second-order correlation measurements). A QD with a band-edge state dipole moment orientation along the longitudinal mode of a gold nanocone is selected for efficient coupling, as shown in Figure 3(c). The dipole moment of the QD is determined by imaging the back focal plane of the fluorescent light. A doughnut-shape emission pattern is attributed to a dipole moment oriented along the longitudinal axis of the nanocone. The localized surface plasmon-polariton resonance of the nanocone is measured by dark-field spectroscopy with p-polarized white light in a different setup. By defocusing the microscope objective, the doughnut shape of the scattered field corresponds to excitation of the longitudinal plasmon mode (Figure 3(c)). The scattered signal is collected and normalized for the uneven white light illumination across the broad wavelength range to precisely determine the localized surface plasmon-polariton resonance

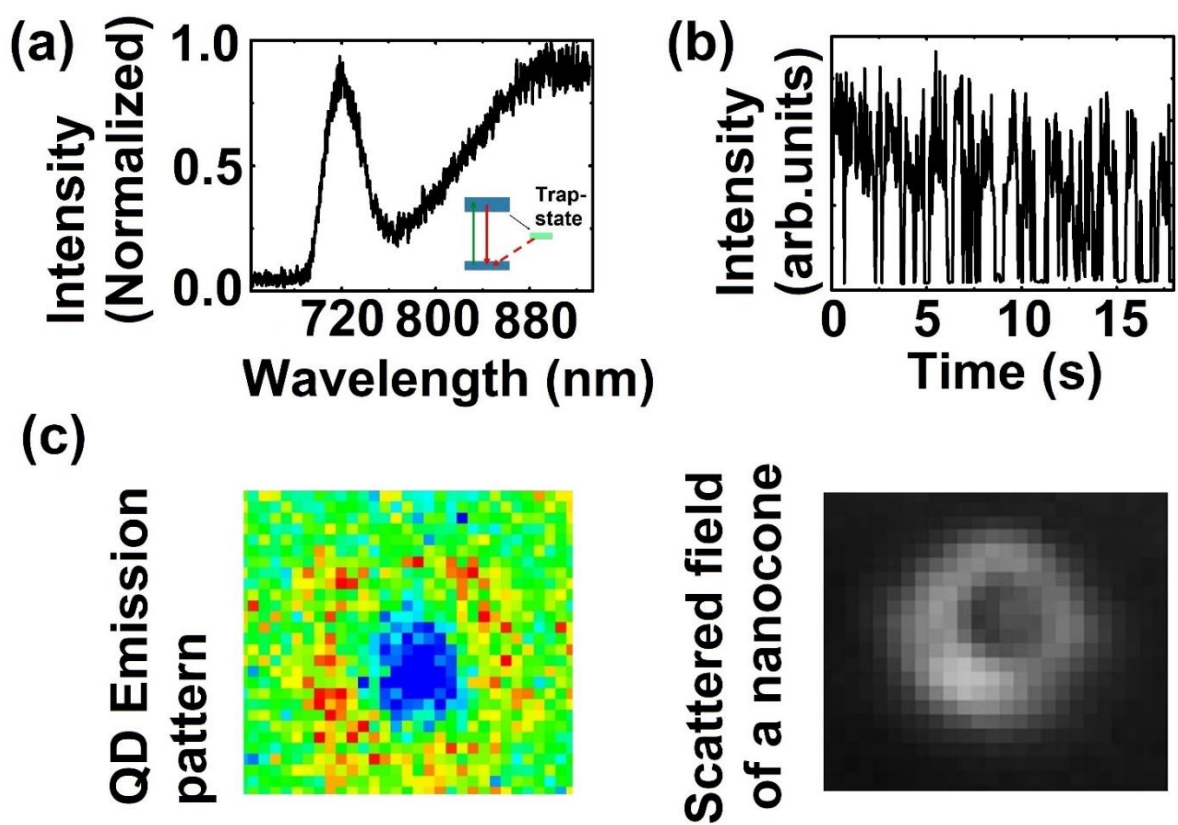

**Figure 3.** (a) Spectrum of a trap-rich single CdTe QD that exhibits both band-edge and surface-state emission. The inset schematically illustrates these transitions. (b) Fluorescence intermittency of a single CdTe QD under continuous excitation. (c) The doughnut shape of the QD emission pattern is attributed to a dipole moment oriented along the longitudinal axis of the nanocone. The localized plasmon resonance of a nanocone is also precisely determined using dark field spectroscopy and back-focal plane imaging.

The excitation laser is detuned from resonance to avoid excitation enhancement (Figure 1(b)). The QD-nanocone distance, polarization of the excitation laser and the dipole orientation of the QD are controlled to modify the coupling strength. Spectrally matching the localized surface plasmon-polariton resonance of a gold nanocone with the band-edge state substantially enhances its emission, while surface-state emission at a wavelength far from the plasmon resonance is significantly suppressed as shown in Figure 4(a) by the red curve. The black curve represents the band-edge and surface-defect state emission of the QD before coupling. When coupling takes place, more than 99% of the emission is associated with the bande-edge transition (red curve). Here, the selective suppression of surface-trap states and the enhancement of the band-edge states is uniquely induced by a plasmonic effect.

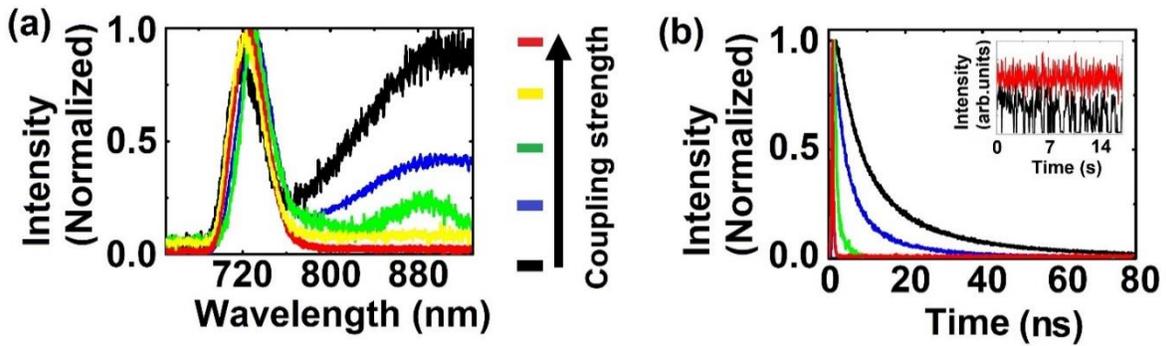

**Figure 4**. (a) The surface defect state of a CdTe QD is modified depending on the coupling strength (QD-nanocone distance and polarization). The QD before coupling (black curve) exhibits substantial defect-state emission. Upon efficient coupling more than 99% of the emission originates from band-edge states (b) Measured excited-state decay curves for different coupling strengths. The extracted lifetime exhibits

a shortening of the spontaneous emission rate by two orders of magnitude (red curve). Inset: fluorescence intermittency of the uncoupled QD (black) is significantly suppressed due to efficient plasmon-coupling (red).

To investigate the modification in the spontaneous emission rate, a 68 ps pulsed laser excites the QD and the emitted photons are detected and histogrammed as a function of delay after the laser pulse. The fluorescence curves are fitted by two decay times. The long lifetime is attributed to the excitonic emission channel and the short lifetime to the biexcitonic emission pathway. The relative weights of each component are defined by the area under the two exponential curves used to fit the data. By controlled positioning of a single QD near the gold nanocone, the spontaneous emission rate of excitons and biexcitons are enhanced by more than two orders of magnitude, as shown in Figure 4(b). The QD before coupling (black curve) exhibits excitonic and biexcitonic lifetimes of 26.50 ns and 6.88 ns, respectively. By bringing the QD in the near field of a gold nanocone, the excitonic and biexcitonic emission rates correspond now to 130 ps and 271 ps, respectively (red curve). The IRF (inset) is deconvoluted from the measured lifetime. Further analysis on the emission dynamics (see supplementary material) shows that the Purcell enhancement of exciton and biexciton transitions amounts to 1460 and 613 folds, respectively. Consequently, the quantum efficiency of the excitonic transition improves by more than a factor of 3. Moreover, Auger non-radiative decay channels are overcome by the faster biexcitonic radiative recombination, leading to a 10-fold enhancement in the quantum efficiency (see supplementary material) turning the defect-rich bare QD into a bright photon source.

In conclusion, we have reported complete suppression of surface-trap emission of a bare CdTe QD solely based on plasmonic coupling. Moreover, we have shown that a strong increase of the radiative excitonic and multiexcitonic transitions (mainly biexciton) suppresses fluorescence intermittency and boost the QD brightness. The biexcitonic enhancement can be mainly attributed to a faster photon emission rate as compared to competitive Auger and other non-radiative rates. The nanoscale size of the hybrid system allows integration into micro- and nano-structures (e.g., microresonators and plannar antennas) for achieving large collection efficiency and directionality [23-25]. The approach can also be used to shape the emission spectrum [26] and for achieving strong coupling at room temperature [27, 28].

**Acknowledgements**

The authors gratefully acknowledge financial support from the University of Siegen, Germany, and the Italian Institute of Technology (IIT), Italy. This article is based upon work from COST Action MP1403 "Nanoscale Quantum Optics," supported by COST (European Cooperation in Science and Technology). A. Flatae would like to thank F. Dinelli for advice on the AFM techniques, M. Ardini and M. Mousavi for advice on functionalization protocols and P. Reuschel for technical assistance.

**Supporting Information**. Brief descriptions of the determination of quantum efficiencies and Purcell enhancements are supplied as Supporting Information.

# Supporting Information

## Determination of quantum efficiencies and Purcell enhancements

Fluorescence lifetime ($\tau$) measurement provides information about the total decay rate as $\tau = 1/\gamma_{total}$. Distinguishing radiative ($\gamma_r$) and nonradiative ($\gamma_{nr}$) decay rates requires information about the quantum efficiency $\eta = \gamma_r/(\gamma_r + \gamma_{nr})$ both in the absence and presence of the nanostructure. Since $\eta$ is very sensitive to the environment, each single quantum dot (QD) can exhibit completely different values. Therefore, the same QD is interrogated during the measurement. The QD before coupling exhibits excitonic and biexcitonic lifetimes of 26.50 ns and 6.88 ns, respectively. By bringing the QD in the near field of a gold nanocone, the excitonic and biexcitonic emission rates now correspond to 130 ps and 271 ps, respectively (as discussed in the manuscript).

The QD count rate is measured using an avalanche photodiode (APD) and the fluorescence signal is imaged using an electron-multiplying charge-coupled device (EMCCD) camera. The signal to noise ratio is determined using spatially- and spectrally-resolved spectroscopy. The pixel array of the EMCCD attached to the spectrometer (Figure S1(a)) allows to quantify the contribution of scatterers and background signal at the single-photon level. In the weak excitation limit, the QD fluorescence signal $S_0$ is enhanced due to the nanocone according to $S = K_{exc}.K_\eta.K_\delta.S_0$, where $K_{exc}$ is the excitation enhancement, $K_\eta$ is the modification of the quantum efficiency and $K_\delta$ refers to the change in the collection efficiency. Note that multiexcitonic emission has to be considered and the parameters are dependent on the dipole orientation and position of the QD with respect to the nanocone.

Since the excitation laser is detuned from the spectral region of QD emission and surface plasmon-polariton resonance, there is no substantial excitation enhancement and $K_{exc}$ is of the order of one. Moreover, a high numerical aperture (NA) oil-immersion microscope objective (Olympus, 60X, 1.42 NA) leads to a large collection efficiency and $K_\delta$ can also be assumed to be close to one. On the other hand, determination of $K_\eta$ and resolution of $\gamma_r$ and $\gamma_{nr}$ for excitonic and biexcitonic emission pathways requires measurements of the second order correlation function and saturation curves, as discussed below.

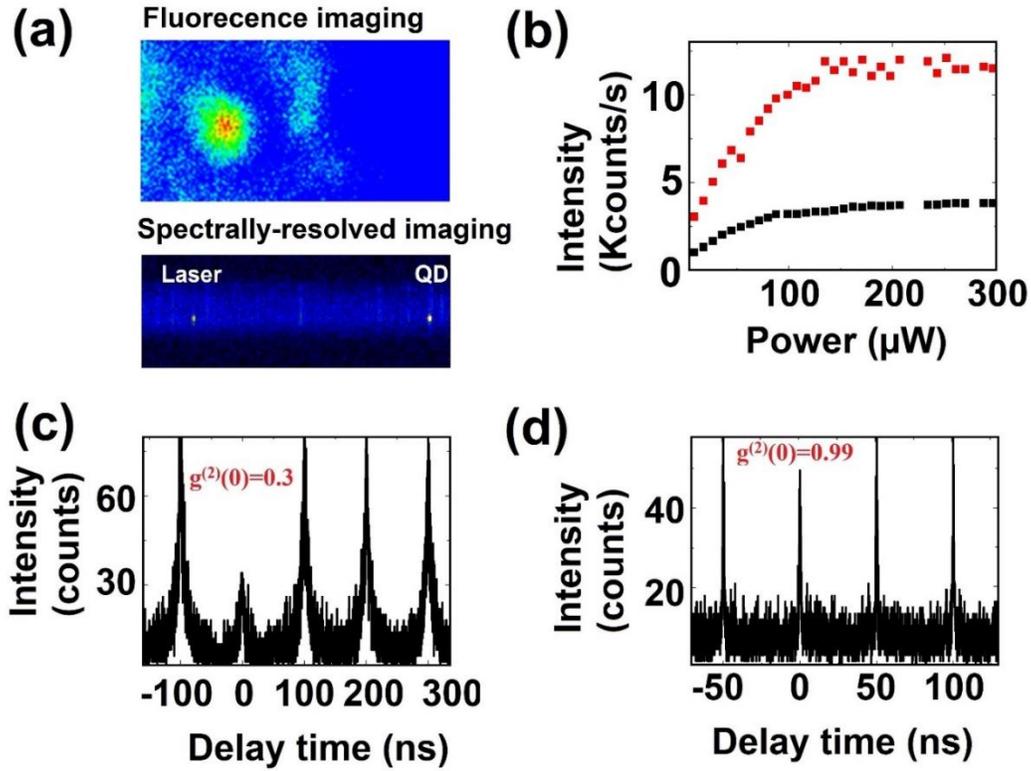

**Figure S1.** (a) The fluorescence emission is imaged using an EMCCD camera and the signal is resolved using the pixel array of the EMCCD camera attached to the spectrometer. One of the laser filters is removed for imaging the laser line. (b) Total fluorescence signal from a QD as a function of the excitation power before coupling (black curve) and after coupling to a nanocone (red curve). (c) Second-order autocorrelation function of the QD before coupling. (d) Second-order autocorrelation function of the QD after coupling to a nanocone.

The quantum efficiency η is the ratio of the number of emitted photons to the number of excitations. Unlike the determination of η for an ensemble of QDs (e.g., using an integrating sphere), precise estimation of the number of absorbed and emitted photons of a single QD is not trivial. In this work we saturate the single QD using a low repetition-rate (1 MHz) laser, so that each incident laser pulse leads to an excitation (each pulse finds the QD in the ground state after the previous excitation). Moreover, this avoids complexity caused by the unavoidable QD dark states. The black curve in Figure S1(b) shows the fluorescence saturation curve as a function of excitation power to quantify the total emission rate of the QD at the tip of the atomic force microscopy (AFM) probe (before coupling). Since the recorded signal has both excitons and biexcitons contribution, the fluorescence decay curve is also recorded at different powers until the contribution from the exciton channel is constant. It is found that beyond 100 µW the excitonic emission is saturated and the fluorescence signal amounts to $S_0$ = 2.8 kcps. Taking into consideration the overall detection efficiency $\beta$ = 2% of our setup, the number of emitted photons can be deduced from $S_0/\beta$. Hence, the excitonic quantum efficiency in the absence of nanocone turns out to be $\eta_0^x \approx$ 14%. This directly tells that the radiative and non-radiative lifetime of the exciton before coupling are $\tau_{0,r}^x$ = 189.3 ns and $\tau_{0,nr}^x$ = 30.8 ns, respectively. The biexcitonic quantum efficiency $\eta_0^{xx}$ is determined using the second-order autocorrelation measurement at zero delay time. Using the Hanbury-Brown Twiss photo autocorrelation measurement, one can use of the relation $g_0^{(2)}(0) = \eta_0^{xx}/\eta_0^x$ to determine the biexcitonic quantum efficiency [1, 2]. The autocorrelation is measured at 10 MHz repetition-rate to decrease the acquisition time required to develop a representative histogram. As shown in Figure 5(c), we find $g_0^{(2)}(0) = 0.3$, corresponding to a biexcitonic quantum efficiency of $\eta_0^{xx} \approx$ 4 %. Hence, the radiative and non-radiative lifetime of the biexciton are $\tau_{0,r}^{xx}$ = 172 ns and $\tau_{0,nr}^{xx}$ = 7.2 ns, respectively.

For the QD positioned at the nanocone tip, the total signal is affected by enhanced radiative contribution from excitons and biexcitons. We find that excitonic saturation occurs at a power of 116 µW, as shown in Figure S1(b) (red curve), and the exciton fluorescence signal corresponds to $S$ = 9.2 kcps. Taking into consideration the same collection efficiency, the excitonic quantum efficiency turns out to be $\eta_c^x$ = 46%, which is more than 3 time higher than the uncoupled QD. This corresponds to an excitonic radiative and non-radiative lifetimes of $\tau_{c,r}^{x} = 0.28$ ns and $\tau_{c,nr}^{x} = 0.24$ ns, respectively. The non-radiative lifetime includes losses due to surface plasmons, which is characterized by antenna efficiency [3]. Taking into consideration $\gamma_{c,nr}^{x} = \gamma_{nr\_plasmon}^{x} + \gamma_{0,nr}^{x}$, where $\gamma_{nr\_plasmon}^{x}$ is the excitonic non-radiative decay rate due to plasmon coupling, the Purcell enhancement factor is $F_r^x = (\gamma_{c,r}^{x} + \gamma_{nr\_plasmon}^{x})/\gamma_{0,r}^{x} \approx 1460$. The biexcitonic quantum efficiency is similarly inferred from the second-order correlation measurement (see Figure S1(d)) and it turns out to be $\eta_c^{xx} \approx 46\%$, which is 10 time higher than the uncoupled QD. Thus, the biexcitonic radiative and non-radiative lifetimes become $\tau_{c,r}^{xx} = 0.59$ ns and $\tau_{c,nr}^{xx} = 0.5$ ns, respectively, during coupling. After similar determination of the bi-excitonic non-radiative decay rate due to plasmon coupling $\gamma_{nr\_plasmon}^{xx}$, the Purcell enhancement factor of excitonic emission is found to be $F_r^{xx} = (\gamma_{c,r}^{xx} + \gamma_{nr\_plasmon}^{xx})/\gamma_{0,r}^{xx} \approx 613$. These results show that efficient coupling to a gold nanocone completely changes the QD photophysics and it turns the defect-rich bare QD into a bright emitter. Finally, we would like to remark that, as expected for large Purcell enhancements, plasmon coupling sets the quantum efficiency of excitonic and biexcitonic transitions to the same value of 46%, which corresponds to the antenna efficiency of the gold nanocone defined as the ratio between the far-field radiated power to the total emitted power [3].